\documentclass[12pt,a4paper]{article}
\usepackage[utf8]{inputenc}
\usepackage[russian]{babel}

\usepackage{amsmath}
\usepackage{amsfonts}
\usepackage{amssymb}
\usepackage{amsthm}
\usepackage{cmap}
\usepackage{enumerate}
\usepackage{indentfirst} 
\usepackage{graphicx}

\usepackage[colorlinks=true,unicode]{hyperref}

\usepackage{fullpage}

\newtheorem{lemma}{Лемма}
\newtheorem{theorem}{Теорема}

\begin{document}

\large

\title{\sc{Анализ переопределенной системы, описывающей специальный класс двумерных движений идеальной жидкости}}

\author{\textbf{Ю.\,В. Шанько\footnote{<<Институт вычислительного моделирования СО РАН>> ФИЦ КНЦ СО РАН; shy70@mail.ru}}}

\date{}
\maketitle

Рассмотрим переопределенную систему уравнений
\begin{gather}
u_t + uu_x + vu_y + p_x = 0, \nonumber \\
v_t + uv_x + vv_y + p_y = 0, \nonumber \\
\label{eq:1}
u_x + v_y = 0, \\
p_t + up_x + vp_y = 0
.
\nonumber
\end{gather}
Здесь $t$ "--- время, $x$, $y$ "--- пространственные координаты, $u$, $v$ "--- компоненты вектора скорости, 
$p$ "--- отклонение давления от заданной величины $p_0$.

Система \eqref{eq:1} является двумерным аналогом общей трехмерной системы,
задача исследования на совместность которой была поставлена
в статье Л.В.~Овсянникова \cite{Ovs99}.
Эта система описывает так называемые тепловые 
(с постоянной плотностью)  движения политропного газа.
К этой же системе сводятся 
изотермические (с постоянной скоростью звука) движения газа при показателе адиабаты не равном $1$.

Анализ системы \eqref{eq:1} 
удобнее выполнять в специальных лагранжевых координатах \cite{NeChup}.
За лагранжеву переменную $\eta$ выбирается давление ($\eta=p$),
а вторая переменная $\xi$ задается так,
чтобы якобиан перехода от $x$, $y$ к $\xi$, $\eta$ равнялся $1$.
Следует отметить, что $\xi$ этим условием определяется неоднозначно.
Полученная система состоит из линейных уравнений 
\begin{gather}
\label{eq:2a}
x_\xi = -y_{tt}, 
\quad 
 y_\xi = x_{tt}
\end{gather}
и нелинейного уравнения
\begin{gather}
\label{eq:2b}
x_\xi y_\eta - x_\eta y_\xi = 1
.
\end{gather}
Заметим, что данный переход к лагранжевым координатам осуществим только при условии $p \neq \mathrm{const}$. 
Движения газа с условием $p = \mathrm{const}$
рассмотрены в работе Л.В.~Овсянникова\cite{Ovs94}.

В работе М.В.~Нещадима и А.П.~Чупахина \cite{NeChup} показано, что общее решение системы \eqref{eq:2a}, \eqref{eq:2b}
не может зависеть от произвольной функции двух переменных
и содержит не более четырех произвольных функций одной переменной.
В статье С.В.~Хабирова \cite{Khabirov} заявлено о построении общего решения аналога системы \eqref{eq:2a}, \eqref{eq:2b}.
Следует отметить, что Теорема 2 работы \cite{Khabirov} является неверной, контрпример к ней построен в последнем разделе данной статьи.
Поэтому утверждение автора \cite{Khabirov} о том, что полученные в его статье формулы задают общее решение \eqref{eq:2a}, \eqref{eq:2b}, осталось недоказанным.
Кроме того, приведенный в указанной работе список точных решений не является полным.
Система эквивалентная \eqref{eq:1} изучалась в статье Н.А.~Иногамова \cite{Inogamov} в связи с исследованиями по лазерному термоядерному синтезу.
В указанной статье автор рассмотрел два класса решений и заявил (без доказательства), что других решений не существует.

В \cite{NeChup} показано, что система \eqref{eq:2a}, \eqref{eq:2b} допускает
группу непрерывных преобразований, порождаемую операторами
\begin{gather}
X_1=\varphi(\eta)\partial_\xi,
\quad
X_2=\partial_\eta,
\quad
X_3=t\partial_x,
\quad
X_4=t\partial_y,
\nonumber \\
X_5=\partial_x,
\quad
X_6=\partial_y,
\quad
X_7=\partial_t,
\quad
X_8=-y\partial_x+x\partial_y,
\label{symmetry}
\\
X_9=t\partial_t+2\xi\partial_\xi-2\eta\partial_\eta,
\quad
X_{10}=x\partial_x+y\partial_y+2\eta\partial_\eta.
\nonumber
\end{gather}

\section{Приведение системы к пассивному виду}

Прежде всего сделаем некоторые замечания.
Все рассмотрение ведется локально, все функции считаются дифференцируемыми нужное число раз.
Объемные вычисления проводились с помощью системы компьютерной алгебры REDUCE \cite{Reduce}.
Под совместностью системы уравнений понимается наличие у нее непустого множества решений.

Введя комплекснозначную функцию $z=x+iy$,
перепишем систему \eqref{eq:2a}, \eqref{eq:2b} следующим образом:
\begin{gather}
\label{eq:z_xi}
z_\xi=iz_{tt}
, \\
\label{eq:z2b}
\frac{i}{2}(z_\xi \overline{z}_\eta - \overline{z}_\xi z_\eta) = 1
\end{gather}
(черта над символом обозначает комплексное сопряжение).

Целью настоящего раздела является приведение системы \eqref{eq:z_xi}, \eqref{eq:z2b} к пассивному
виду \cite{Finikov}.

Вначале, вместо системы \eqref{eq:z_xi}, \eqref{eq:z2b} будет удобно рассматривать векторную систему,
которая является ее следствием. 
Введем в двумерном векторном пространстве билинейную кососимметрическую форму.
Для векторов $\mathbf{a}=(a^1,a^2)$ и $\mathbf{b}=(b^1,b^2)$ положим
\[
\mathbf{a} \vee \mathbf{b}
=\begin{vmatrix} a^1 & b^1 \\ a^2 & b^2 \end{vmatrix}
.
\]
Из  \eqref{eq:z_xi}, \eqref{eq:z2b} следует, что
вектор 
$\mathbf{z}
=(z^1,z^2)
=(z_\xi,z_\eta)$ удовлетворяет уравнениям:
\begin{gather}
\label{eq:vecz_xi}
\mathbf{z}_\xi=i\mathbf{z}_{tt}
, \\
\label{eq:vec2b}
\frac{i}{2}
\,
\mathbf{z} \vee \overline{\mathbf{z}}
 = 1
.
\end{gather}

Положим
\begin{gather*}
\alpha=\frac{i}{2}\, \mathbf{z}\vee\mathbf{\overline{z}}
, \\
\beta=\frac{1}{2}\left( \mathbf{z}_t\vee\mathbf{\overline{z}} - \mathbf{z}\vee\mathbf{\overline{z}}_t \right)
, \\
\gamma=-\frac{i}{2}\left( \mathbf{z}_{tt}\vee\mathbf{\overline{z}} - 2 \,\mathbf{z}_t\vee\mathbf{\overline{z}}_t + \mathbf{z}\vee\mathbf{\overline{z}}_{tt} \right)
, \\
\delta=-\frac{1}{2}\left( \mathbf{z}_{ttt}\vee\mathbf{\overline{z}} -3\,\mathbf{z}_{tt}\vee\mathbf{\overline{z}}_t 
      +3\,\mathbf{z}_t\vee\mathbf{\overline{z}}_{tt} -\mathbf{z}\vee\mathbf{\overline{z}}_{ttt} \right)
, \\
\varepsilon=\frac{i}{2}\left(\mathbf{z}_{tttt}\vee\mathbf{\overline{z}} - 4 \,\mathbf{z}_{ttt}\vee\mathbf{\overline{z}}_t 
+6\, \mathbf{z}_{tt}\vee\mathbf{\overline{z}}_{tt}
-4\,\mathbf{z}_t\vee\mathbf{\overline{z}}_{ttt} + \mathbf{z}\vee\mathbf{\overline{z}}_{tttt} \right)
.
\end{gather*}
Функции $\alpha$, $\beta$, $\gamma$, $\delta$, $\varepsilon$ выбраны так,
что они являются вещественными и 
в силу уравнения \eqref{eq:vecz_xi} справедливы соотношения:
\begin{gather}
\alpha_\xi+\beta_t=\beta_\xi+\gamma_t=\gamma_\xi+\delta_t=\delta_\xi+\varepsilon_t=0
.
\label{eq:6}
\end{gather}

\begin{lemma}
\label{lem:1}
Положим
\begin{gather*}
\Delta_1 = -4(\alpha \alpha_{tt} - \alpha_t^2 + \alpha \gamma -\beta^2)
,\\
\Delta_2 = 2(\alpha \alpha_{ttt} - \alpha_t \alpha_{tt} + \alpha \gamma_t + \gamma \alpha_t - 2 \beta \beta_t)
,\\
\Delta_3 = 2( \beta \alpha_{tt} - 2 \alpha_t \beta_t + \alpha \beta_{tt} + \alpha \delta - \beta \gamma )
,\\
\Delta_4 =  -\alpha_t \alpha_{ttt} + \alpha_{tt}^2 - \alpha_t \gamma_t + \beta \beta_{tt} + \beta \delta - \gamma^2 
,\\
\Delta_5 =  -\beta \alpha_{ttt} + 2 \beta_t \alpha_{tt} - \alpha_t \beta_{tt} - \alpha_t \delta + 2 \gamma \beta_t - \beta \gamma_t 
.
\end{gather*}
Тогда справедливо тождество
\begin{gather}
\label{eq:lin_gen}
\Delta_1 \mathbf{z}_{tt} + (\Delta_2 + i \Delta_3) \mathbf{z}_t + (\Delta_4 + i \Delta_5) \mathbf{z} = 0
.
\end{gather}
Кроме того, неравенство $\Delta_1\neq 0$ эквивалентно условию
\begin{gather}
\label{neq:1}
\mathbf{z}_t\vee\mathbf{z} \neq 0
.
\end{gather} 
\end{lemma}

\begin{proof}
Рассмотрим определитель
\begin{gather*}
\label{eq:det5}
\Delta=
\begin{vmatrix}
m_1 & m_2 & m_3 & m_4 & m_5 \\
z^1_{tt} & z^1_t &  iz^1_t & z^1 &  iz^1 \\
z^2_{tt} & z^2_t &  iz^2_t & z^2 &  iz^2 \\
\overline{z^1_{tt}} & \overline{z^1_t} & -i\overline{z^1_t} & \overline{z^1} & -i\overline{z^1} \\
\overline{z^2_{tt}} & \overline{z^2_t} & -i\overline{z^2_t} & \overline{z^2} & -i\overline{z^2} \\
\end{vmatrix}
.
\end{gather*}
Раскроем $\Delta$ по первой строке.
Функции $\Delta_j$ подобраны так, чтобы они были равны соответствующим
алгебраическим дополнениям
(в этом можно убедиться непосредственным вычислением).
Получим тождество:
\begin{gather*}
\Delta = \Delta_1m_1 + \Delta_2m_2 + \Delta_3m_3 + \Delta_4m_4 + \Delta_5m_5
.
\end{gather*}
Если подставить в определитель $\Delta$ вместо первой строки вторую или третью,
то он, очевидно, занулится.
Получаем соотношения
\begin{gather*}
\Delta_1 z^j_{tt} + (\Delta_2 + i \Delta_3) z^j_t + (\Delta_4 + i \Delta_5) z^j = 0
,
\end{gather*}
из которых следует первое утверждение леммы.

Второе утверждение леммы следует из цепочки равенств:
\begin{gather*}
\label{eq:m1}
\Delta_1=
\begin{vmatrix}
z^1_t &  iz^1_t & z^1 &  iz^1 \\
z^2_t &  iz^2_t & z^2 &  iz^2 \\
\overline{z^1_t} & -i\overline{z^1_t} & \overline{z^1} & -i\overline{z^1} \\
\overline{z^2_t} & -i\overline{z^2_t} & \overline{z^2} & -i\overline{z^2} \\
\end{vmatrix}
=4
\begin{vmatrix}
z^1_t & z^1\\
z^2_t & z^2\\
\end{vmatrix}
\begin{vmatrix}
\overline{z^1_t}& \overline{z^1}\\
\overline{z^2_t}& \overline{z^2}\\
\end{vmatrix}
=
4(\mathbf{z}_t\vee\mathbf{z}) (\overline{ \mathbf{z}_t \vee \mathbf{z} })
.
\end{gather*}
\end{proof}

Из леммы следует, что
при выполнении условия \eqref{neq:1}
и заданных функциях $\Delta_j$
вектор-функция $\mathbf{z}$ удовлетворяет линейному уравнению
второго порядка
\begin{gather}
\label{eq:vecz_t2}
\mathbf{z}_{tt}=2iK\mathbf{z}+T\mathbf{z}
,
\end{gather}
с коэффициентами
\begin{gather*}
K=\frac{i\Delta_2 -\Delta_3}{2\Delta_1}
,\quad
T=\frac{-\Delta_4 -i \Delta_5}{\Delta_1}
.
\end{gather*}

Рассмотрим вначале случай, когда условие \eqref{neq:1} не выполняется,
т.е, когда $\mathbf{z}_t\vee\mathbf{z}=0$.
Это означает, что векторы $\mathbf{z}_t$ и $\mathbf{z}$ линейно зависимы:
\begin{gather}
\label{eq:deg1}
\mathbf{z}_t=S\mathbf{z}
.
\end{gather}
Здесь $S=S(t, \xi, \eta)$ "--- некоторая комплекснозначная функция.

\begin{lemma}
\label{lem:deg}
Для совместности системы уравнений \eqref{eq:vecz_xi}, \eqref{eq:vec2b}, \eqref{eq:deg1}
необходимо выполнение условий
\begin{gather*}
S+\overline{S}=S_t=S_\xi=0
.
\end{gather*}
\end{lemma}
\begin{proof}
Продифференцируем \eqref{eq:vec2b} по $t$  в  силу уравнения \eqref{eq:deg1}:
\begin{gather*}
-\frac{1}{4}(S + \overline{S})\mathbf{z}\vee\mathbf{\overline{z}}=0
.
\end{gather*}
Отсюда
\begin{gather}
\label{eq:deg2}
\overline{S} = -S
.
\end{gather}
Подставим $\mathbf{z}_t$ из \eqref{eq:deg1} в уравнение \eqref{eq:vecz_xi}:
\begin{gather}
\label{eq:deg3}
\mathbf{z}_\xi=
i(S_t+S^2)\mathbf{z}
.
\end{gather}
Из условия совместности уравнений
\eqref{eq:deg1} и \eqref{eq:deg3}
следует, что
\begin{gather}
\label{eq:deg5}
S_\xi=i(S_t+S^2)_t
.
\end{gather}
Продифференцируем \eqref{eq:vec2b} по $\xi$ в силу уравнения \eqref{eq:deg3}:
\begin{gather*}
\frac{1}{2}(\overline{S}_t+\overline{S}^2-S_t-S^2)\mathbf{z}\vee\mathbf{\overline{z}}=0
.
\end{gather*}
Отсюда, с учетом \eqref{eq:deg2}, получаем $S_t=0$,
тогда из \eqref{eq:deg5} следует, что и $S_\xi=0$.
\end{proof}

Из Леммы \eqref{lem:deg} следует, что $S=iN$, где
$N(\eta)$ "---  некоторая вещественная функция.
Запишем уравнение \eqref{eq:deg1} в координатах, с учетом \eqref{eq:z_xi}:
\begin{gather}
\label{eq:deg7}
z_{ttt}=iNz_{tt}
, \\
\label{eq:deg8}
z_{t\eta}=iNz_\eta
.
\end{gather}
Выписывая условие совместности этой системы, получаем 
$N_\eta=0$, иными словами, $N=\mathrm{const}$.
В дальнейшем будет показано, что \eqref{eq:deg7}, \eqref{eq:deg8} можно рассматривать как частный случай некоторых более общих уравнений.

Перейдем теперь к рассмотрению случая  $\mathbf{z}_t\vee\mathbf{z} \neq 0$.
\begin{lemma}
\label{lem:identity}
Справедливо тождество
\begin{gather}
\Omega=
(\alpha_{tttt}+2 \gamma_{tt}+\varepsilon)   (\beta^2 -\alpha (\gamma +\alpha_{tt}) +\alpha_t^2)
+(\alpha_{tt}+\gamma)(  4 \beta_t^2 +  (\alpha_{tt}-\gamma)^2 ) 
+ \nonumber \\
\label{eq:5}
+\alpha (\beta_{tt}+\delta)^2
+\alpha (\alpha_{ttt}+\gamma_t)^2
+2 (\beta_{tt}+\delta) (\alpha_{tt} \beta - 2 \alpha_t \beta_t - \beta \gamma)
+\\
+2 (\alpha_{ttt}+\gamma_t)   (\alpha_t \gamma - \alpha_{tt} \alpha_t - 2 \beta_t \beta)
=0
.
\nonumber
\end{gather}
\end{lemma}
\begin{proof}
Непосредственным вычислением можно проверить справедливость равенств:
\begin{gather*}
\Omega=8i
\begin{vmatrix}
-2i \alpha & -\beta -i \alpha_t & \frac{i}{2} (\gamma-\alpha_{tt}) - \beta_t \\
\beta -i \alpha_t & \frac{-i}{2} (\gamma+\alpha_{tt}) & \frac{-i}{4} (\gamma+\alpha_{tt})_t - \frac{1}{4}(\delta+\beta_{tt}) \\
\frac{i}{2} (\gamma-\alpha_{tt}) + \beta_t & \frac{-i}{4} (\gamma+\alpha_{tt})_t + \frac{1}{4}(\delta+\beta_{tt}) & \frac{-i}{8} (\alpha_{tttt}+2\gamma_{tt}+\varepsilon)
\end{vmatrix}
= \\ =
8i
\begin{vmatrix}
\mathbf{z}\vee\mathbf{\overline{z}} & \mathbf{z}\vee\mathbf{\overline{z}}_t & \mathbf{z}\vee\mathbf{\overline{z}}_{tt} \\
\mathbf{z}_t\vee\mathbf{\overline{z}} & \mathbf{z}_t\vee\mathbf{\overline{z}}_t & \mathbf{z}_t\vee\mathbf{\overline{z}}_{tt} \\
\mathbf{z}_{tt}\vee\mathbf{\overline{z}} & \mathbf{z}_{tt}\vee\mathbf{\overline{z}}_t & \mathbf{z}_{tt}\vee\mathbf{\overline{z}}_{tt}
\end{vmatrix}
.
\end{gather*}
Последний определитель равен нулю, поскольку каждый его столбец является линейной комбинацией двух столбцов
$(z^1,z^1_t,z^1_{tt})$
и
$(z^2,z^2_t,z^2_{tt})$.
\end{proof}

Уравнение \eqref{eq:vec2b} дает нам $\alpha=1$. Тогда, дифференцируя уравнения \eqref{eq:6} нужное число раз по $t$,
получаем:
\begin{gather}
\beta_t=\gamma_{tt}=\delta_{ttt}=\varepsilon_{tttt}=0
.
\label{eq:7}
\end{gather}
Подстановка этих соотношений в 
\eqref{eq:5} дает уравнение
\begin{gather}
(\beta^2 - \gamma) \varepsilon  
+\delta^2
+\gamma_t^2
-2 \beta \gamma \delta 
+\gamma^3 
=0
.
\label{eq:8}
\end{gather}

\begin{lemma}
\label{lem:gamma_t}
Если выполнены уравнения \eqref{eq:7}, \eqref{eq:8}, 
то
$\gamma_t=0$. 
\end{lemma}

\begin{proof}
Обозначим $\gamma_t=\mu$. Функция $\gamma$ линейна по $t$, поэтому $\mu$ от $t$ не зависит.
Предположим, что $\gamma_t=\mu \neq 0$,
тогда, найдутся такие $\xi_0$, $\eta_0$, что $\mu(\xi_0,\eta_0)\neq 0$.
Возьмем $t_0$ такое, что $\gamma(t_0,\xi_0,\eta_0)=\beta^2(\xi_0,\eta_0)$.
Рассмотрим уравнение \eqref{eq:8} в точке $(t_0,\xi_0,\eta_0)$:
\[
\mu^2(\xi_0,\eta_0)
+(\beta^3(\xi_0,\eta_0)-\delta(t_0,\xi_0,\eta_0))^2
=0
.
\]
Функции $\mu$, $\beta$, $\delta$ вещественные, поэтому $\mu(\xi_0,\eta_0)=0$.
Противоречие.
\end{proof}
Из доказанной леммы и одного из уравнений \eqref{eq:6}, а именно, $\gamma_\xi+\delta_t=0$ следует, что $\delta_{tt}=0$.
Значит вместо \eqref{eq:7} мы можем записать следующие соотношения:
\begin{gather}
\label{eq:7:1}
\beta_t=\gamma_t=\delta_{tt}=0
.
\end{gather}
Используя \eqref{eq:7:1}, выпишем формулы для коэффициентов $K$ и $T$ из уравнения \eqref{eq:vecz_t2}:
\begin{gather*}
K=\frac{\delta - \beta \gamma}{4 (\gamma - \beta^2)}
, \quad
T=\frac{\beta \delta - \gamma^2}{4 (\gamma - \beta^2)}
.
\end{gather*}

Так как, по предположению, $\mathbf{z}_t\vee\mathbf{z} \neq 0$,
то из Леммы \ref*{lem:1} следует,
что $\Delta_1 \neq 0$, а значит 
\begin{gather}
\label{neq:0}
\gamma-\beta^2 \neq 0
.
\end{gather}
Из той же леммы получаем, что $\mathbf{z}$ удовлетворяет уравнению второго порядка \eqref{eq:vecz_t2}.
Перейдем от векторной записи уравнений к скалярной.
Положим 
$\mathbf{z}=(z_\xi, z_\eta)$
и заменим в полученных уравнениях производные $z_\xi$  в силу \eqref{eq:z_xi}.
Уравнения \eqref{eq:lin_gen}, \eqref{eq:z2b} и условие \eqref{neq:1} запишутся следующим образом:
\begin{gather}
\label{eq:z_t4}
z_{tttt} =  2 i K z_{ttt}   + T z_{tt}
, \\
\label{eq:z_t2eta}
z_{tt\eta} = 2 i K z_{t\eta} + T z_\eta
,
\\
\label{eq:detz}
-\frac{1}{2}(z_{tt}\overline{z}_\eta+\overline{z}_{tt}z_\eta)=1
,
\\
\label{neq:2}
z_{ttt}z_\eta - z_{tt}z_{t\eta} \neq 0
.
\end{gather}

\begin{lemma}
\label{lem:delta_t}
При  выполнении условия \eqref{neq:0}
система, состоящая из уравнений \eqref{eq:z_t4}, \eqref{eq:z_t2eta}, \eqref{eq:detz}, может быть совместна, только если 
$\delta_t=0$. 
\end{lemma}

\begin{proof}
Введем обозначения:
$c=\gamma-\beta^2$,
$d=(\delta-2\beta\gamma+\beta^3)/c^2$,
$f=\beta\beta_\eta d_t - cd_{t\eta}$,
$g=dd_{t\eta}-d_td_\eta$,
$h=2\beta c d-\beta^2-3c$,
$w=ic^2d_tz_\eta-c_\eta z_{tt}$.

Из уравнений \eqref{eq:7:1} следует, что
\[
c_t=d_{tt}=f_t=g_t=0
.
\]
Из условия \eqref{neq:0} получаем, что $c\neq 0$.
Предположим, что 
$d_t=\delta_t/c^2\neq 0$
и покажем, что в этом случае рассматриваемая система уравнений не имеет решений.
Запишем уравнения \eqref{eq:z_t4}, \eqref{eq:z_t2eta} в новых обозначениях:
\begin{gather}
\label{eq:z_t4:1}
z_{tttt} = \frac{i}{2}(d + \beta) z_{ttt} +  \frac{c}{4}(\beta d - 1) z_{tt}
, \\
\label{eq:z_t2eta:1}
z_{tt\eta} = \frac{i}{2}(d + \beta) z_{t\eta} +  \frac{c}{4}(\beta d - 1) z_\eta
.
\end{gather}

Выпишем условие совместности этих уравнений:
\begin{gather}
\label{eq:comp1}
2id w_t + (\beta d - 1) w - 2 i (\beta_\eta + d_\eta c) z_{ttt}  - c(\beta d )_\eta  z_{tt} = 0
.
\end{gather}

Заметим, что в силу  \eqref{eq:z_t4:1}, \eqref{eq:z_t2eta:1} функция $w$ удовлетворяет уравнению
\begin{gather}
\label{eq:w_t2}
w_{tt} = \frac{i}{2}(d + \beta) w_t +  \frac{c}{4}(\beta d - 1) w
.
\end{gather}

Продифференцировав \eqref{eq:comp1} по $t$ в силу \eqref{eq:z_t4:1}, \eqref{eq:w_t2},
придем к уравнению
\begin{gather}
\label{eq:comp1t}
2 (2i d_t - c d^2  - 1) w_t + (2 \beta d_t + i cd (\beta d - 1)) w
+ \\ \nonumber
+2 (\beta \beta_\eta + c^2 d d_\eta - 2i c d_{t\eta}) z_{ttt}
- c ( 2 (\beta d_t)_\eta + i (\beta_\eta + cd_\eta) (\beta d - 1) ) z_{tt} 
= 0
.
\end{gather}
Повторив эту операцию еще дважды, получим еще два линейных
дифференциальных уравнения относительно функций $z$ и $w$.
Вместе с \eqref{eq:comp1}, \eqref{eq:comp1t} они образуют
систему линейных однородных алгебраических уравнений относительно $z_{ttt}$, $z_{tt}$, $w_t$, $w$.
Эта система должна иметь нетривиальное решение, так как равенство нулю $z_{tt}$ противоречит
\eqref{eq:detz}.
Следовательно, определитель системы должен равняться нулю.
Приравняв к нулю вещественную и мнимую части определителя, получим:
\begin{gather}
\label{eq:det_re}
\beta_\eta^2 d_t^2  (36 d_t^2 c^2 + h^2) 
+ c \beta_\eta g d_t ( c^2 d^2 h - c h - h^2 - 12 d_t^2 c^2) 
- \\
-  2 c d f h \beta_\eta d_t
+ c^4 g^2 + f(f + \beta c g) (h - c) +c^3 f g d ( 2  - \beta d) = 0
\nonumber
,
\\
\label{eq:det_im}
3 c^3 d_t \beta_\eta g d^2
- 2 c(
3 \beta_\eta f d_t + 2 \beta_\eta g d_t \beta c + f g c) d
+ \\
+ 4 f^2 - 2 \beta_\eta f d_t \beta + 3 \beta_\eta g d_t c^2  + 3 f g \beta c - g^2 c^3 = 0
.
\nonumber
\end{gather}
Левая часть уравнения \eqref{eq:det_im} "--- многочлен второй степени относительно $d$.
Нетрудно проверить, что коэффициенты многочлена от $t$ не зависят,
а $d$ существенно зависит от $t$, поскольку,
по предположению, $d_t\neq 0$.
Приравнивая к нулю эти коэффициенты,
получим уравнения:
\begin{gather}
\label{eq:det_im2}
\beta_\eta g = 0
,\\
\label{eq:det_im1}
3 \beta_\eta f d_t + 2 \beta_\eta g d_t \beta c + f g c = 0
, \\
\label{eq:det_im0}
4 f^2 - 2 \beta_\eta f d_t \beta + 3 \beta_\eta g d_t c^2  + 3 f g \beta c - g^2 c^3 = 0
.
\end{gather}

Покажем, что $\beta_\eta=0$.
Предположим, что это не так, тогда из \eqref{eq:det_im2} следует, что $g=0$, а из \eqref{eq:det_im1},
что и $f=0$.
Подставив $f=g=0$ в 
\eqref{eq:det_re}, получим
\begin{gather*}
\beta_\eta^2 d_t^2  (36 d_t^2 c^2 + h^2) = 0
.
\end{gather*}
Очевидно, что при сделанных предположениях
левая часть этого уравнения не может обратиться в ноль.

Итак, $\beta_\eta=0$.
Уравнения \eqref{eq:det_im1}, \eqref{eq:det_im0} запишутся следующим образом:
\begin{gather*}
fg= 0
, \\
4 f^2 + 3 f g \beta c - g^2 c^3 = 0
.
\end{gather*}
Отсюда, в силу того, что $c \neq 0$, получаем $f = g = 0$.
Так как, по определению, 
$f=\beta\beta_\eta d_t - cd_{t\eta}$,
$g=dd_{t\eta}-d_td_\eta$
и, как только что было показано, $\beta_\eta=0$,
то и $d_\eta=0$.
Подставив эти соотношения в 
\eqref{eq:comp1}, \eqref{eq:comp1t},
получим линейную алгебраическую систему относительно $w_t$, $w$:
\begin{gather*}
2id w_t + (\beta d - 1) w = 0
, \\
2 (2i d_t - c d^2  - 1) w_t + (2 \beta d_t + i cd (\beta d - 1)) w = 0
.
\end{gather*}
Определитель системы равен
$4id_t+2(\beta d - 1)$
и отличен от нуля, так как, по предположению, $d_t\neq 0$.
Следовательно $w=0$, 
откуда
\begin{gather*}
	z_\eta=-i\frac{c_\eta}{c^2 d_t}z_{tt}
	.
\end{gather*}
Подставив $z_\eta$ в уравнение \eqref{eq:detz}, мы занулим его левую часть, а его правая часть равна $1$.
Противоречие.
\end{proof}

Поскольку $\beta_t=\gamma_t=\delta_t=0$, то из формул для $K$ и $T$ следует, что $K_t=T_t=0$.
Выпишем теперь условия совместности \eqref{eq:vecz_xi} и \eqref{eq:vecz_t2}:
\begin{gather*}
2iK_\xi \mathbf{z}_t + T_\xi\mathbf{z} = 0
.
\end{gather*}
Отсюда следует, что $K_\xi=T_\xi=0$, так как в противном случае
$\mathbf{z}_t\vee\mathbf{z} = 0$.

Рассмотрим теперь отдельно линейные уравнения \eqref{eq:z_t4}, \eqref{eq:z_t2eta}.
\begin{lemma}
\label{lem:eta_dependance}
Пусть $z_{tt}\neq 0$, выполнено условие \eqref{neq:2},  
функции $K$ и $T$ не зависят от $t$ и $\xi$ и
система уравнений \eqref{eq:z_t4}, \eqref{eq:z_t2eta} совместна.
Тогда при $K_\eta\neq 0$ каждое ее решение удовлетворяет уравнениям
\begin{gather}
\label{eq:z_t3}
z_{ttt} = iN z_{tt}
,
\\
\label{eq:z_teta}
z_{t\eta}=iNz_\eta-iN^{-2}N_\eta z_{tt}
.
\end{gather}
где $N$ "--- некоторая вещественная функция от $\eta$.
Если же функция $K$ является константой, то и $T$ должна быть константой.
\end{lemma}
\begin{proof}
Выпишем условия совместности уравнений \eqref{eq:z_t4}, \eqref{eq:z_t2eta}:
\begin{gather}
\label{eq:comp3}
2K_\eta z_{ttt} - iT_\eta z_{tt}=0
. 
\end{gather}
Если $K_\eta = 0$, то, так как $z_{tt} \neq 0$, то и $T_\eta = 0$, иными словами, $K$ и $T$ "--- константы.

Если же $K_\eta\neq 0$, то уравнение \eqref{eq:comp3} можно записать в виде \eqref{eq:z_t3},
причем
\begin{gather}
\label{eq:T_eta}
T_\eta=2NK_\eta
.
\end{gather}
Выпишем условие совместности \eqref{eq:z_t3} и
\eqref{eq:z_t4}
(при $z_{tt}\neq 0$):
\begin{gather}
\label{eq:T}
T = 2KN-N^2
.
\end{gather}
Условие совместности \eqref{eq:z_t3} и
\eqref{eq:z_t2eta},
с учетом \eqref{eq:T},
запишется так:
\begin{gather}
\label{eq:comp5}
(2K-N)^2(z_{t\eta}-iNz_\eta)+iN_\eta z_{tt}=0
.
\end{gather}
Из уравнения \eqref{eq:z_t3} и условия \eqref{neq:2} следует, что 
\begin{gather}
\label{neq:3}
z_{t\eta} \neq iN z_\eta
.
\end{gather}
Подставим $T$ из \eqref{eq:T} в уравнение \eqref{eq:T_eta}:
\begin{gather*}
2(K-N)N_\eta=0
.
\end{gather*}
Покажем, что $K=N$.
Если это не так, то тогда $N_\eta=0$ и
из уравнения \eqref{eq:comp5} и условия \eqref{neq:3} следует, что
$K=N/2$.
Но тогда, очевидно,
$K$ является константой,
что противоречит нашему предположению.
Таким образом, $K=N$.
Подставив $K$ 
в \eqref{eq:comp5}
и разрешив его относительно $z_{t\eta}$, получим уравнение \eqref{eq:z_teta}.
\end{proof}

Случай, когда $K$ и $T$ являются константами, требует дальнейшего рассмотрения.
Положим $K=k$, $T=k^2+m$,
где $k,\, m \in \mathbb{R}$.
Уравнения \eqref{eq:z_t4}, \eqref{eq:z_t2eta}  запишутся следующим образом:
\begin{gather}
\label{eq:z_t4:2}
z_{tttt} = 2ik z_{ttt} + (k^2+m)z_{tt}
, \\
\label{eq:z_t2eta:2}
z_{tt\eta} = 2ik z_{t\eta} + (k^2+m)z_\eta
.
\end{gather}
Продифференцируем \eqref{eq:detz} два раза по $t$ в силу уравнений
\eqref{eq:z_t4:2}, \eqref{eq:z_t2eta:2}:
\begin{gather}
\nonumber
ik(z_{tt}\overline{z}_{t\eta}-\overline{z}_{tt}z_{t\eta}+\overline{z}_{ttt}z_\eta-z_{ttt}\overline{z}_\eta)
-(z_{ttt}\overline{z}_{t\eta}+\overline{z}_{ttt}z_{t\eta})-
\\ -
\label{eq:z_teta:1}
(k^2+m)(z_{tt}\overline{z}_\eta+\overline{z}_{tt}z_\eta)
=0
.
\end{gather}
Выразим $\overline{z}_\eta$ из уравнения \eqref{eq:detz}:
\begin{gather}
\label{eq:conjz_eta}
\overline{z}_\eta = \frac{-z_\eta\overline{z}_{tt}-2}{z_{tt}}
.
\end{gather}
Подставим найденное $\overline{z}_\eta$ в \eqref{eq:z_teta:1}:
\begin{gather}
\nonumber
i(i z_{tt}\overline{z}_{ttt} - i \overline{z}_{tt}z_{ttt} - 2k z_{tt}\overline{z}_{tt} )
(z_{tt}z_{t\eta} - z_\eta z_{ttt}) -
\\ - 
\label{eq:zteta_new}
2(z_{ttt}^2 - 2ikz_{tt}z_{ttt} - (k^2 + m)z_{tt}^2) = 0
.
\end{gather}

Предположим, что
\begin{gather}
\label{neq:4}
i z_{tt}\overline{z}_{ttt} - i \overline{z}_{tt}z_{ttt} - 2k z_{tt}\overline{z}_{tt} \neq 0
,
\end{gather} 
тогда из уравнения \eqref{eq:zteta_new} можно выразить $z_{t\eta}$:
\begin{gather}
\label{eq:z_teta:2}
z_{t\eta}=\frac{z_\eta z_{ttt}}{z_{tt}} - 2i
\frac{z_{ttt}^2 - 2ikz_{tt}z_{ttt} - (k^2+m)z_{tt}^2}
{z_{tt}(iz_{tt}\overline{z}_{ttt} - i\overline{z}_{tt}z_{ttt} - 2k z_{tt}\overline{z}_{tt} )}
.
\end{gather}
Если же условие \eqref{neq:4} не выполняется, то из \eqref{eq:zteta_new} получаем уравнения:
\begin{gather}
\label{eq:indeterm:1}
\overline{z}_{tt}z_{ttt} - z_{tt}\overline{z}_{ttt} - 2ik z_{tt}\overline{z}_{tt} = 0
, \\ \label{eq:indeterm:2}
-z_{ttt}^2 + 2ikz_{tt}z_{ttt} + (k^2 + m)z_{tt}^2 = 0
.
\end{gather}

\begin{lemma}
\label{lem:indeterm}
Пусть $z_{tt}\neq 0$ и выполнено условие \eqref{neq:2},
тогда система уравнений \eqref{eq:z_t2eta:2}, \eqref{eq:indeterm:1}, \eqref{eq:indeterm:2}
совместна только при $k=m=0$.
\end{lemma}

\begin{proof}
Из \eqref{eq:indeterm:2} следует, что 
\begin{gather}
\label{eq:indeterm:3}
z_{ttt}=lz_{tt}
,
\end{gather}
где $l$ "--- 
корень квадратного уравнения
\begin{gather}
\label{eq:indeterm:4}
l^2 - 2ikl-k^2-m=0
.
\end{gather}
Подставив \eqref{eq:indeterm:3} в \eqref{eq:indeterm:1}, придем к соотношению
\begin{gather}
\label{eq:indeterm:5}
2ik = l-\overline{l}
,
\end{gather}
Подставив $2ik$ из \eqref{eq:indeterm:5} в \eqref{eq:indeterm:4}, получим $k^2+m=l\overline{l}$.
Используя найденные соотношения между константами, перепишем уравнение \eqref{eq:z_t2eta:2} в следующем виде:
\begin{gather}
\label{eq:indeterm:6}
z_{tt\eta} = (l-\overline{l})z_{t\eta} + l\overline{l}z_\eta
.
\end{gather}
Выписав условие совместности уравнений \eqref{eq:indeterm:3} и \eqref{eq:indeterm:6},
получим
\begin{gather*}
\label{eq:indeterm:7}
\overline{l}^2(z_{t\eta}-lz_\eta)=0
.
\end{gather*}
Из \eqref{eq:indeterm:3} и условия \eqref{neq:2} следует, что $z_{t\eta}\neq lz_\eta$, поэтому $l=0$.
Тогда из \eqref{eq:indeterm:5} и \eqref{eq:indeterm:4} получаем утверждение леммы.
\end{proof}

Подставив $k=m=0$ в \eqref{eq:z_t2eta:2}, \eqref{eq:indeterm:2}, получим уравнения
\begin{gather}
\label{eq:z_t2eta:3}
z_{tt\eta}=0
, \\
z_{ttt}=0
\label{eq:z_t3:1}
.
\end{gather}

Подведем итоги исследования на совместность.
Прежде всего заметим, что
уравнения \eqref{eq:deg7}, \eqref{eq:deg8} при $N\neq 0$
являются частным случаем уравнений \eqref{eq:z_t3}, \eqref{eq:z_teta}.
Если же $N=0$, то, очевидно, любое решение \eqref{eq:deg7}, \eqref{eq:deg8} будет решением уравнений \eqref{eq:z_t2eta:3}, \eqref{eq:z_t3:1}.

Непосредственным вычислением можно показать,
что в каждом из рассмотренных случаев все условия совместности будут следствиями уже выписанных уравнений.
Это позволяет нам говорить о пассивности полученных систем.
Сформулируем результат в виде теоремы.
Поскольку в предыдущих рассуждениях под следствиями уравнений мы понимали не только
уравнения, полученные дифференцированием исходных, но и уравнения, полученные комплексным сопряжением,
то, чтобы не отступать от классического определения пассивной системы \cite{Finikov},
при формулировке теоремы следует добавить к некоторым уравнениям комплексно-сопряженные к ним.

\begin{theorem}
Множество решений системы уравнений \eqref{eq:z_xi}, \eqref{eq:z2b}
является объединением множеств решений следующих пассивных
систем:
\begin{enumerate}[(A)]
\item \label{sys:Deg}
система состоящая из уравнений: \eqref{eq:z_xi} и сопряженного к нему, \eqref{eq:conjz_eta}, \eqref{eq:z_t2eta:3}, \eqref{eq:z_t3:1} и сопряженного к нему;
\item \label{sys:ArbFun}
система состоящая из уравнений:  \eqref{eq:z_xi} и сопряженного к нему, \eqref{eq:z_t3} и сопряженного к нему, \eqref{eq:z_teta}, \eqref{eq:conjz_eta}
"--- с произвольной функцией 
$N(\eta)$ не равной тождественно нулю;
\item \label{sys:Standard}
система состоящая из уравнений:  \eqref{eq:z_xi} и сопряженного к нему, \eqref{eq:z_t4:2} и сопряженного к нему, \eqref{eq:conjz_eta}, \eqref{eq:z_teta:2}
"--- с произвольными константами $k$ и $m$.
\end{enumerate}
\end{theorem}

Следует отметить, что три полученные системы можно преобразовать к эквивалентным пассивным \textit{ортономным} системам \cite{Finikov}.
Однако это сильно усложнит формулы. Вместо этого, будут приведены решения полученных систем.

\section{Свойства решений}

В этом разделе рассматриваются свойства решений системы \eqref{eq:1} с точки зрения теории движения идеальной несжимаемой жидкости.

\textbf{1.}
В гидродинамике система \eqref{eq:1} задает двумерные движения  идеальной жидкости 
с дополнительным условием постоянства давления в частице.
Это условие позволяет интерпретировать каждое решение \eqref{eq:1}, как движение жидкости со свободной границей,
определяемой соотношением 
$p=0$.

\textbf{2.} Несложно проверить, что завихренность 
\[ \omega =  
x_t \vee x + y_t  \vee y
= \beta . \]
Из \eqref{eq:6}, \eqref{eq:7} и Леммы \ref{lem:gamma_t} следует, что $\beta_t=\beta_\xi=0$,
т.е. $\beta$ может зависеть только от $\eta$.
Иными словами завихренность $\omega=\beta$ и давление $p=\eta$
связаны функциональным соотношением $\omega=\omega(p)$.

\textbf{3.} Получим условия, при которых жидкость ограничена движущейся твердой стенкой.
Рассмотрим двумерное движение жидкости, при котором она ограничена некоторой (вообще говоря) движущейся кривой.
Будем считать, что в лагранжевых координатах кривая задается уравнением $\eta=\mathrm{const}$.

Как известно \cite{Rashevsk50},
форма плоской кривой полностью определяется, с точностью до выбора начальной точки отчета дуги,
по ее кривизне как функции от натурального параметра $s$.
Следовательно, кривизна кривой $\kappa$, как функция от $s$ и времени $t$ должна иметь вид:
\begin{gather}
\label{eq:curv:1}
\kappa=\nu(s-s_0(t))
.
\end{gather}
Здесь $s_0(t)$ задает выбор начальной точки отчета в зависимости от момента времени,
а функция $\nu$ определяет форму кривой.
Нетрудно проверить, что все такие $\kappa$ удовлетворяют уравнению
\begin{gather}
\label{eq:curv:2}
\kappa_t\kappa_{ss}-\kappa_s\kappa_{ts}=0
.
\end{gather}
С другой стороны, в общее решение
\eqref{eq:curv:2} кроме функций вида \eqref{eq:curv:1} также входят функции $\kappa=\kappa(t)$.
Поэтому, условие представимости $\kappa$ в виде \eqref{eq:curv:1} 
эквивалентно тому, что выполняется \eqref{eq:curv:2} и $\kappa_s\neq 0$, либо  $\kappa_s=\kappa_t=0$.

Вычислим теперь кривизну кривой для рассматриваемой системы.
В~силу уравнения \eqref{eq:z_xi} она задается формулой
\[
\kappa=\frac{x_\xi y_{\xi\xi} - y_\xi x_{\xi\xi}}{(x_\xi^2+y_\xi^2)^{3/2}}
=\frac{i(z_\xi\overline{z}_{\xi\xi}-\overline{z}_\xi z_{\xi\xi}) }{2(z_\xi\overline{z}_\xi)^{3/2}}
=\frac{z_{tt}\overline{z}_{tttt}+\overline{z}_{tt} z_{tttt} }{2(z_{tt}\overline{z}_{tt})^{3/2}}
.
\] 
Переход от производных по натуральному параметру $s$ к производным по $\xi$ осуществляется по формуле: 
\[
\kappa_s=\dfrac{\kappa_\xi}{s_\xi}
,
\]
где
\[
s_\xi=\sqrt{x_\xi^2+y_\xi^2}=\sqrt{z_\xi \overline{z}_\xi}=|z_\xi|
.
\]
Поэтому, условие \eqref{eq:curv:2} эквивалентно следующему:
\begin{gather}
\label{eq:curv:3}
\kappa_t   \left(\frac{\kappa_\xi}{|z_\xi|}\right)_\xi -
\kappa_\xi \left(\frac{\kappa_\xi}{|z_\xi|}\right)_t   = 0
.
\end{gather}

Определим теперь в каких случаях
кривые $\eta=\mathrm{const}$
на решениях системы \eqref{eq:z_xi}, \eqref{eq:z2b}
можно рассматривать в качестве движущихся твердых стенок.

Для системы 
\ref{sys:Deg}
кривизна $\kappa$ равна нулю,
а для системы 
\ref{sys:ArbFun}
\[
\kappa
=-N^2(\eta)|z_{tt}|^{-1}
.
\]
В обоих случаях $\kappa_s=\kappa_t=0$.

Для системы \ref{sys:Standard} кривизна задается формулой:
\begin{gather*}
\kappa=\dfrac{ik(\overline{z}_{tt}z_{ttt}-z_{tt}\overline{z}_{ttt}) +(k^2+m)z_{tt}\overline{z}_{tt} }{(z_{tt} \overline{z}_{tt})^{3/2}}
.
\end{gather*}
Непосредственным вычислением можно проверить, что $\kappa$ удовлетворяет уравнению \eqref{eq:curv:3},
а также  получить следующее соотношение:
\begin{gather*}
\kappa_s
=\dfrac{-2k}{|z_{tt}|}\kappa_t
.
\end{gather*}
Поэтому, при $k\neq 0$ производная $\kappa_s$ обращается в нуль одновременно с~$\kappa_t$.
Если же $k=0$, то очевидно, $\kappa_s=0$. Вычислим
\begin{gather*}
\kappa_t
=\frac{-m(z_{tt}\overline{z}_{ttt} + \overline{z}_{tt} z_{ttt} )}{2(z_{tt} \overline{z}_{tt})^{3/2}}
=-m\frac{{|z_{tt}|}_t}{|z_{tt}|^2}
.
\end{gather*}
Итак, форма кривой сохраняется, кроме случая $k=0$, $m\neq 0$.

Таким образом, для всех решений \eqref{eq:z_xi}, \eqref{eq:z2b}
(исключая решения системы 
\ref{sys:Standard}
при $k=0$, $m\neq 0$)
форма кривой $\eta=\mathrm{const}$ не меняется со временем.
Это означает, что любую из этих кривых можно рассматривать в качестве движущейся твердой стенки.

\bigskip

Для того, чтобы жидкость была ограничена только кривыми $\eta=p=\mathrm{const}$,
эти кривые не должны иметь самопересечений.
Это достигается не всегда.

\section{Точные решения}

Перейдем к построению точных решений.
Решения приводятся с точностью до действия группы $G$, 
порождаемой операторами \eqref{symmetry}
и дискретными преобразованиями
обращения времени $t \to -t$ и
отражения $\xi \to -\xi$, $y \to -y$.

Решение системы \ref{sys:Deg} дается формулой:
\begin{gather*}
z=\xi+S(\eta)t+i\left(\eta-\frac{t^2}{2}\right)
,
\end{gather*}
где $S$ "--- произвольная гладкая функция.

Решение системы \ref{sys:ArbFun}:
\begin{gather*}
z=S(\eta )e^{i(N(\eta )t-N^2(\eta )\xi)}
, \\
S(\eta )S'(\eta )N^2(\eta )=1
\end{gather*}

Систему \ref{sys:Standard} можно значительно упростить с помощью 
подходящего преобразования из группы $G$
и введения новой независимой переменной.
Положим
\begin{gather*}
J=(z_{ttt} - ikz_{tt})\overline{z}_\eta + (\overline{z}_{ttt} + ik\overline{z}_{tt})z_\eta
\end{gather*}
и обозначим левую часть неравенства \eqref{neq:4} следующим образом:
\begin{gather*}
I=i (z_{tt}\overline{z}_{ttt} - \overline{z}_{tt}z_{ttt}) - 2k z_{tt}\overline{z}_{tt}
.
\end{gather*}

Можно проверить, что $J_t=J_\xi=I_t=I_\xi=0$ в силу уравнений рассматриваемой системы.
Продолжим оператор $X_1$ на $J$:
\begin{gather*}
X_1=\varphi\partial_\xi-I \varphi' \partial_J
.
\end{gather*}
В силу условия \eqref{neq:4} $I\neq 0$.
Следовательно, выбрав подходящую функцию $\varphi$, можно занулить $J$ действием оператора $X_1$.
Из уравнений \eqref{eq:conjz_eta} и $J=0$ следует, что
\begin{gather*}
z_\eta=2(iz_{ttt}+kz_{tt})I^{-1}
,\\
\overline{z}_\eta=2(-i\overline{z}_{ttt}+k\overline{z}_{tt})I^{-1}
.
\end{gather*}
Введем независимую переменную (новую лагранжеву координату) $\sigma$,
связанную с $\eta$ соотношением:
\begin{gather}
\label{eq:Std:eta}
\eta_\sigma= I=  i(z_{tt}\overline{z}_{ttt} - \overline{z}_{tt}z_{ttt} ) - 2kz_{tt}\overline{z}_{tt}
.
\end{gather}
Получим систему:
\begin{gather}
z_{tttt} = 2ikz_{ttt} + (k^2 + m)z_{tt}, \nonumber \\
\label{eq:Std:lin}
z_\xi = iz_{tt}, \\
z_\sigma = 2(iz_{ttt} + kz_{tt}) \nonumber
.
\end{gather}
Таким образом, нелинейная система \ref{sys:Standard} сведена к
линейной системе \eqref{eq:Std:lin}
и нелинейному уравнению для определения $\eta$ \eqref{eq:Std:eta}.
Линейная однородная система \eqref{eq:Std:lin} обладает конечномерным пространством решений.
Поэтому для нее можно выписать фундаментальную систему решений (ФСР),
которая состоит из четырех комплекснозначных функций.
Две из этих функций, а именно, $z_1=1$ и $z_2=t$ не зависят от того, какие значения констант $k$ и $m$ мы выберем.
Вид еще двух функций $z_3$ и $z_4$ определяется корнями характеристического уравнения
\[
\lambda^4-2ik\lambda^3-(k^2+m)\lambda^2=0
,
\]
которое строится по первому уравнению \eqref{eq:Std:lin}.

Далее рассматриваются только те решения системы \ref{sys:Standard}, которые
удовлетворяют условиям \eqref{neq:1} и \eqref{neq:4},
т.е. не являются решениями двух предыдущих систем.

\textbf{1.} В случае $m=k=0$ в
ФСР входят функции: $z_3=t^3/6+it\xi+2i\sigma$, $z_4=t^2/2+i\xi$.
Преобразованиями из группы $G$ решения \eqref{eq:Std:eta}, \eqref{eq:Std:lin} можно свести к виду:
\begin{gather*}
z=t^3/6 - \xi + i(t^2/2 + t\xi + 2\sigma)
, \\
\eta=-2\sigma
.
\end{gather*}

\textbf{2.} При $m=-k^2$ в
ФСР входят функции: $z_3=\exp( 2ikt-4ik^2\xi+8k^3\sigma)$, $z_4=t^2/2+i\xi+2k\sigma$.
Преобразованиями из $G$ решения системы \eqref{eq:Std:eta}, \eqref{eq:Std:lin} можно свести к виду ($k=1/2$):
\begin{gather*}
z=\exp( it-i\xi+\sigma) - \xi + i(t^2/2+\sigma)
, \\
\eta=\exp(2\sigma)/2 - \sigma
.
\end{gather*}

Данное решение отличается от решения, задающего трохоидальные волны Герстнера только слагаемым 
$t^2/2$.
Это слагаемое возникает вследствие того, что волны Герстнера описывают движение жидкости в постоянном поле тяжести,
а в нашем случае внешние силы равны нулю.

\textbf{3.} Рассмотрим случай $m<0$, $m+k^2\neq 0$. Положим $k=(a+b)/2$, $m=-(a-b)^2/4$.
В ФСР входят функции: $z_3=\exp( iat-ia^2\xi+a^2(a-b)\sigma)$, 
$z_4=\exp( ibt-ib^2\xi+b^2(b-a)\sigma)$.
Преобразованиями из группы $G$ решения \eqref{eq:Std:eta}, \eqref{eq:Std:lin} можно свести к виду
($m=-1$, $k \ge 0$, $k \neq 1$):
\begin{gather*}
z=\exp( i(k+1)t-(k+1)^2(i\xi-2\sigma))+ 
\\+
\exp( i(k-1)t-(k-1)^2(i\xi+2\sigma))
, \\
\eta=((k+1)^2\exp(4(k+1)^2\sigma) + (k-1)^2\exp(-4(k-1)^2\sigma))/2
.
\end{gather*}

В работе \cite{AbrYak} течения такого типа названы {\em птолемеевскими}.

\textbf{4.} При $m=0$
в ФСР входят функции:
$z_3=\exp( ikt-ik^2\xi)$,
$z_4=(t-2k\xi-2ik^2\sigma)  \exp( ikt-ik^2\xi)$.
Преобразованиями из $G$ решения системы \eqref{eq:Std:eta}, \eqref{eq:Std:lin} можно свести к виду ($k=1$):
\begin{gather*}
z=(t-2\xi-2i\sigma) \exp( it-i\xi)
, \\
\eta=2(\sigma^2+2\sigma)
.
\end{gather*}

\textbf{5.} В случае $m>0$, $m\neq k^2$
в ФСР входят функции:
$z_{3;4}=\exp( i(k\pm i\sqrt{m})t-i(k\pm i\sqrt{m})^2 (\xi\mp 2\sigma\sqrt{m}))$.
Преобразованиями из группы $G$ решения \eqref{eq:Std:eta}, \eqref{eq:Std:lin} можно свести к виду:
\begin{gather*}
z=\exp( ie^{\theta i}t-ie^{2\theta i} (\xi-2\sigma\sin\theta))+
\exp( ie^{-\theta i}t-ie^{-2\theta i}(\xi+2\sigma\sin\theta))
, \\
\eta=\exp(-4\sigma\sin\theta\sin 2\theta) \cos(2\theta+4\sigma\sin\theta\cos 2\theta)
,
\end{gather*}
где $\theta$ "--- некоторая константа такая, что  $\sin\theta\cos 2\theta\neq 0$;
$k=\cos\theta$, $m=\sin^2\theta$.

\textbf{6.} При $m=k^2\neq 0$
в ФСР входят функции:
$z_{3;4}=\exp( k(i\pm 1)t - 2k^2 (\pm \xi + 2k\sigma))$.
Преобразованиями из $G$ решения системы \eqref{eq:Std:eta}, \eqref{eq:Std:lin} можно свести к виду ($m=k=1$):
\begin{gather*}
z 
= \exp( i(t-\theta) - 4\sigma - t + 2\xi)+
  \exp( i(t+\theta) - 4\sigma + t - 2\xi)
, \\
\eta=2\exp(-8\sigma)\sin 2\theta
.
\end{gather*}
Здесь $\theta$ "--- произвольная константа такая, что
$\sin 2\theta \neq 0$.

\section{Замечания к статье С.В.~Хабирова}
В работе С.В.~Хабирова \cite{Khabirov} рассматривается система аналогичная \eqref{eq:2a}, \eqref{eq:2b}.
При исследовании ее на совместность вводятся 
два бесконечных набора функций $p_k$, $q_k$ ($k=0,1,2,\ldots$),
которые зависят от времени $t$ и лагранжевых переменных $i$ и $j$.
В \cite{Khabirov} показано, что эти функции удовлетворяют системе уравнений:
\begin{gather}
\nonumber
p_0=-1,
\quad
4p_{k+1}'=p_k'''+(q_k)_j,
\quad
q_k'=(p_k)_j,
\\ \label{eq:khabirov1}
p_{k+1}(4p_kp_{k-1}
-(p_{k-1}')^2
-q_{k-1}^2)
-p_{k-1}({p_k'}^2+q_k^2)+
\\ \nonumber +
(p_k'p_{k-1}'-q_kq_{k-1})(p_{k-1}''-2p_k)-
p_k(p_{k-1}''-2p_k)^2+
\\ + \nonumber
q_{k-1}'(p_k'q_{k-1}+q_kp_{k-1}'-p_kq_{k-1}')
=0
.
\end{gather}
Здесь штрих означает производную по $t$, а нижний индекс $j$ "--- по соответствующей переменной
(в последнем уравнении \eqref{eq:khabirov1} исправлены допущенные в~\cite{Khabirov} опечатки).

В Теореме 2 рассматриваемой работы утверждается, что все решения \eqref{eq:khabirov1} зависят только от $i$, иными словами, не зависят от $t$ и $j$.
Приведем контрпример к теореме.
Зададим функции $p_k$, $q_k$ рекуррентными соотношениями:
\begin{gather}
p_0=-1,
\quad
q_0=0,
\nonumber
\\
\label{eq:khabirov2}
p_{k+1}=\frac{{p_k'}^2+q_k^2+h_k s j^{-2k-1} }{4p_k}
,\\
q_{k+1}=\frac{-2p_k''q_k+2p_k'q_k'+4p_{k+1}q_k - h_k t s j^{-2k-2}}{4p_k}
,
\nonumber
\end{gather}
где $s$ "--- произвольная функция от лагранжевой переменной $i$, а
\begin{gather*}
h_k=2^{-4k}\prod_{n=0}^{k-1}(s^2+n^2)
.
\end{gather*}
В частности, получим:
\begin{gather*}
p_1=\frac{-s}{4j},
\quad
q_1=\frac{ts}{4j^2},
\\
p_2=\frac{-js^2-t^2s}{16j^3},
\quad
q_2=\frac{2tjs^2+t^3s}{16j^4}
.
\end{gather*}
Построенные по рекуррентным формулам функции $p_k$, $q_k$
удовлетворяют уравнениям \eqref{eq:khabirov1}.
Для того, чтобы доказать это,
присоединим к \eqref{eq:khabirov1}
следующие уравнения:
\begin{gather}
\label{eq:khabirov5}
q_k'=-\frac{tp_k'+2kp_k}{2j}
,\\ \nonumber
p_k''=\frac{j{p_k'}^2-sp_k^2+tp_kq_k+jq_k^2+sj^{-2k}h_k }{2jp_k}
.
\end{gather}
После этого непосредственной подстановкой функций из \eqref{eq:khabirov2} в
\eqref{eq:khabirov1} и \eqref{eq:khabirov5} проверяется обращение этих уравнений в тождества
при начальных значениях $k$,
а также обосновывается, что эти уравнения будут выполнены при любом натуральном
$k$, в силу справедливости уравнений при меньших $k$,
иными словами, применяется 
метод математической индукции.

Рассмотрим теперь решения, полученные в \cite{Khabirov}.
Запишем их в наших обозначениях:

\textbf{1.}
$z=(t^2-\eta/2)-(t\alpha(\eta)-2\xi)i$;

\textbf{2.}
$z=(t^3+2\xi)-(t^2-6t\xi-\eta/2)i$;

\textbf{3.}
$z=\beta(\eta)\exp( i(-\xi \alpha^2(\eta)+t\alpha(\eta)))$,
$\alpha^2\beta\beta_\eta=1$;

\textbf{4.}
$z=t^2+2\ln|\delta|+2\xi i+\delta\exp( i(t-\xi))$,
$\eta=\delta^2/2-4\ln |\delta|$;

\textbf{5.}
$z=(1+A\tg\varphi -i(t+2\xi+A) ) \exp( i(-\xi-t))$,
\newline
$\eta=3A\tg\varphi+A^2(1+\tg^2\varphi)/2$, 
$A\neq 0$;

\textbf{6.}
$z=(m+1-i(t+2\xi) )\exp( i(-\xi-t))$,
$\eta=3m+m^2/2$;

\textbf{7.}
$z=-iv\exp( i(-\lambda^2\xi+\lambda t))-iN|v|^{-1/\lambda^2}\exp( i(-\xi+\varphi\pm t))$,
\newline
$2\eta=\lambda^2 v^2+N^2 |v|^{-2/\lambda^2}$;

\textbf{8.}
$z=-i\exp(-2\xi+t+i t)-\eta\exp(2\xi-t+i(t+\varphi))/(2\cos\varphi)$.

В таблице \ref{tab:1} приводится соответствие между решениями из \cite{Khabirov} и решениями из данной статьи.
\newline 
\begin{table}
\caption{Соответствие между решениями}
\label{tab:1}
\begin{tabular}{|c|c|c|c|c|c|c|c|c|}
\hline \rule[-2ex]{0pt}{5.5ex} Номер решения & 1 & 2 & 3 & 4 & 5 & 6 & 7 & 8 \\ 
\hline \rule[-2ex]{0pt}{5.5ex} Номер формулы из \cite{Khabirov} & (5.4) & (5.5) & (6.2) & (6.4) & (7.4) & (7.5) & (8.2) & (9.4) \\ 
\hline \rule[-2ex]{0pt}{5.5ex} Система & A & C.1 & B & C.2 & C.4 & C.4 & C.3 & C.6 \\ 
\hline 
\end{tabular} 
\end{table}

Решение \textbf{5} можно свести к решению \textbf{6}.
Действительно, положим  в формулах \textbf{5} $A\tg\varphi=m$. Получим 
\begin{gather*}
z=(1+m -i(t+2\xi+A) ) \exp( i(-\xi-t))
,\\
\eta=3m+(A^2+m^2)/2
.
\end{gather*}
После чего уберем все вхождения константы $A$ в формулах с помощью сдвигов,
порождаемых операторами $X_1$, $X_2$, $X_7$.

Константа $N$ в решении \textbf{7} несущественна, ее можно взять равной $\pm 1$.
Для этого положим $v= N^{\lambda^2/(1+\lambda^2)}v_*$ и выполним растяжение,
порождаемое оператором $X_{10}$.

Решения системы \ref{sys:Standard}.5 в работе \cite{Khabirov} отсутствует.

\bigskip

Автор благодарит О.В~Капцова за советы и внимание к работе.

\end{document}